\documentclass[journal=nalefd,manuscript=article]{achemso}

\usepackage{chemformula} % Formula subscripts using \ch{}
\usepackage[T1]{fontenc} % Use modern font encodings

\author{Chao Lei}
\altaffiliation{Contributed equally to this work}
\email{leichao.ph@gmail.com}
\affiliation[The University of Texas at Austin]
{Department of Physics, The University of Texas at Austin, Austin, Texas 78712,USA}
\author{Bheema L. Chittari}
\altaffiliation{Contributed equally to this work}
\affiliation[Indian Institute of Science Education and Research Kolkata]
{Department of Physical Sciences, Indian Institute of Science Education and Research Kolkata, Mohanpur 741246, West Bengal, India }
\affiliation[University of Seoul]
{Department of Physics, University of Seoul, Seoul 02504, Korea}
\author{Kentaro Nomura}
\affiliation[ Tohoku University]
{Institute for Materials Research, Tohoku University, Sendai Aoba-ku 980-8577, Japan}
\author{Nepal Banerjee}
\affiliation[University of Seoul]
{Department of Physics, University of Seoul, Seoul 02504, Korea}
\altaffiliation{Department of Smart Cities, University of Seoul, Seoul 02504, Korea}
\author{ Jeil Jung}
\email{jeiljung@uos.ac.kr}
\affiliation[University of Seoul]
{Department of Physics, University of Seoul, Seoul 02504, Korea}
\altaffiliation{Department of Smart Cities, University of Seoul, Seoul 02504, Korea}
\author{Allan H. MacDonald}
\affiliation[The University of Texas at Austin]
{Department of Physics, The University of Texas at Austin, Austin, Texas 78712,USA}

\title[An \textsf{achemso} demo]
  {Magnetoelectric Response of Antiferromagnetic $\rm CrI_3$ Bilayers}


\usepackage{graphicx}
\usepackage{mathrsfs}
\usepackage{bm}
\usepackage{hyperref}
\usepackage{dcolumn}
\usepackage{amsmath}
\usepackage{amssymb}
\usepackage{color}
\usepackage{caption}
\usepackage[normalem]{ulem}
\newcommand{\bk}{{\mathbf k}}
\newcommand{\bS}{{\mathbf S}}
\newcommand{\bL}{{\mathbf L}}

\newcommand{\bR}{{\mathbf R}}
\newcommand{\bRp}{{\mathbf R^{\prime} }}

\newcommand{\bohr}{\mu_{_B}}
\DeclareRobustCommand{\rchi}{{\mathpalette\irchi\relax}}
\newcommand{\irchi}[2]{\raisebox{\depth}{$#1\chi$}} % inner command, used by \rchi

\hypersetup{colorlinks=true, linkcolor=black, citecolor=black, urlcolor=black  }
\RequirePackage[normalem]{ulem} %DIF PREAMBLE
\RequirePackage{color}\definecolor{RED}{rgb}{1,0,0}\definecolor{BLUE}{rgb}{0,0,1} %DIF PREAMBLE
\begin{document}

%%%%%%%%%%%%%%%%%%%%%%%%%%%%%%%%%%%%%%%%%%%%%%%%%%%%%%%%%%%%%%%%%%%%%
%% The "tocentry" environment can be used to create an entry for the
%% graphical table of contents. It is given here as some journals
%% require that it is printed as part of the abstract page. It will
%% be automatically moved as appropriate.
%%%%%%%%%%%%%%%%%%%%%%%%%%%%%%%%%%%%%%%%%%%%%%%%%%%%%%%%%%%%%%%%%%%%%
%\begin{tocentry}

%\end{tocentry}

%%%%%%%%%%%%%%%%%%%%%%%%%%%%%%%%%%%%%%%%%%%%%%%%%%%%%%%%%%%%%%%%%%%%%
%% The abstract environment will automatically gobble the contents
%% if an abstract is not used by the target journal.
%%%%%%%%%%%%%%%%%%%%%%%%%%%%%%%%%%%%%%%%%%%%%%%%%%%%%%%%%%%%%%%%%%%%%
\begin{abstract}
We predict that layer antiferromagnetic bilayers formed from van der Waals (vdW) materials
with weak inter-layer versus intra-layer exchange coupling have strong magnetoelectric response that 
can be detected in dual gated devices where internal displacement fields and carrier densities can be varied independently.
We illustrate this strong temperature dependent magnetoelectric response in bilayer CrI$_3$ at charge neutrality
by calculating the gate voltage dependent total magnetization 
through Monte Carlo simulations and mean-field 
solutions of the anisotropic Heisenberg model informed from density functional theory and experimental data, and present a simple model for electrical control of magnetism by electrostatic doping.
\end{abstract}

%%%%%%%%%%%%%%%%%%%%%%%%%%%%%%%%%%%%%%%%%%%%%%%%%%%%%%%%%%%%%%%%%%%%%
%% Start the main part of the manuscript here.
%%%%%%%%%%%%%%%%%%%%%%%%%%%%%%%%%%%%%%%%%%%%%%%%%%%%%%%%%%%%%%%%%%%%%
\section{Introduction}
Spintronics studies the interplay between electrical and magnetic properties of materials
and underlies an important technology that was based so far mostly on the
properties~\cite{Wang2011,Maruyama2009,Wu2010,He2010} of ferromagnetic metals.
There has long been interest in expanding spintronics to
semiconductors\cite{Spintronics_Review}, which tend to have properties that are more subject to electrical control~\cite{Matsukura2015}, and usually have antiferromagnetic order. 
Antiferromagnets do have some potential advantages for spintronics, which have attracted attention recently.~\cite{Jungwirth2018,Baltz2018,Jungwirth2016} 
These advantages include insensitivity to magnetic fields, absence of stray fields,
the possibility of terahertz manipulation for ultrafast switching,
multiple stable domain configurations.
Here, we consider the new class of two-dimensional van der Waals magnetic materials 
that have been fabricated for the first time relatively recently~\cite{Huang2017,Gong2017}, 
for which tunneling magnetoresistance~\cite{Song2018,Wang2018,Kim2018,Klein_2018}, electrical
control of magnetic configurations by doping~\cite{Jiang2018a,Jiang2018b,Huang2018}, and 
pressure-induced interlayer magnetic transition\cite{Li_2019} have already been demonstrated in CrI$_3$ bilayers.

Surprisingly, CrI$_3$ bilayers have antiferromagnetic interlayer interactions~\cite{Huang2017,Jiang2018a,Jiang2018b,Huang2018}
even though bulk CrI$_3$ is ferromagnetic.
It has thus been widely studied to understand the magnetism of CrI$_3$ bilayers~\cite{FernandezRossier,Sivadas2018,Jiang2018,Jang2018,Chen2019,Guo_2019,Soriano_2020,Thiel_2019,Sun_2019,Zhang2019,McCreary_2020,Ubrig_2019,Klein_2019}.
Since the bilayers are coupled via van der Waals (vdW) forces,
the interlayer magnetic exchange interactions are also weak,
and it was shown that the interlayer magnetic exchange can be tuned from antiferromagnetic to ferromagnetic just by electrostatic doping~\cite{Jiang2018a,Jiang2018b,Huang2018}.
Electric fields in van der Waals bilayers also alter the magnon dispersion~\cite{Pershoguba2018},
opening a gap between bands associated with the two spins per unit cell in the two-dimensional
honeycomb lattices and opening the possibility of topological magnon bands
when the electric field breaks the inversion symmetry and generates
Dzyaloshinskii-Moriya interactions~\cite{Chisnell2015,Owerre2017,Liu2018_DMI}.
Even though applications await further development, still in progress~\cite{Bonilla2018,Hara2018,roomtemp,CrTe2},
of well-controlled magnetic van der Waals bilayers that order at room temperature,
it is not premature to contemplate the particular advantages of this class of antiferromagnetic semiconductors.

\begin{figure}[htp]
\includegraphics[width=0.9\linewidth]{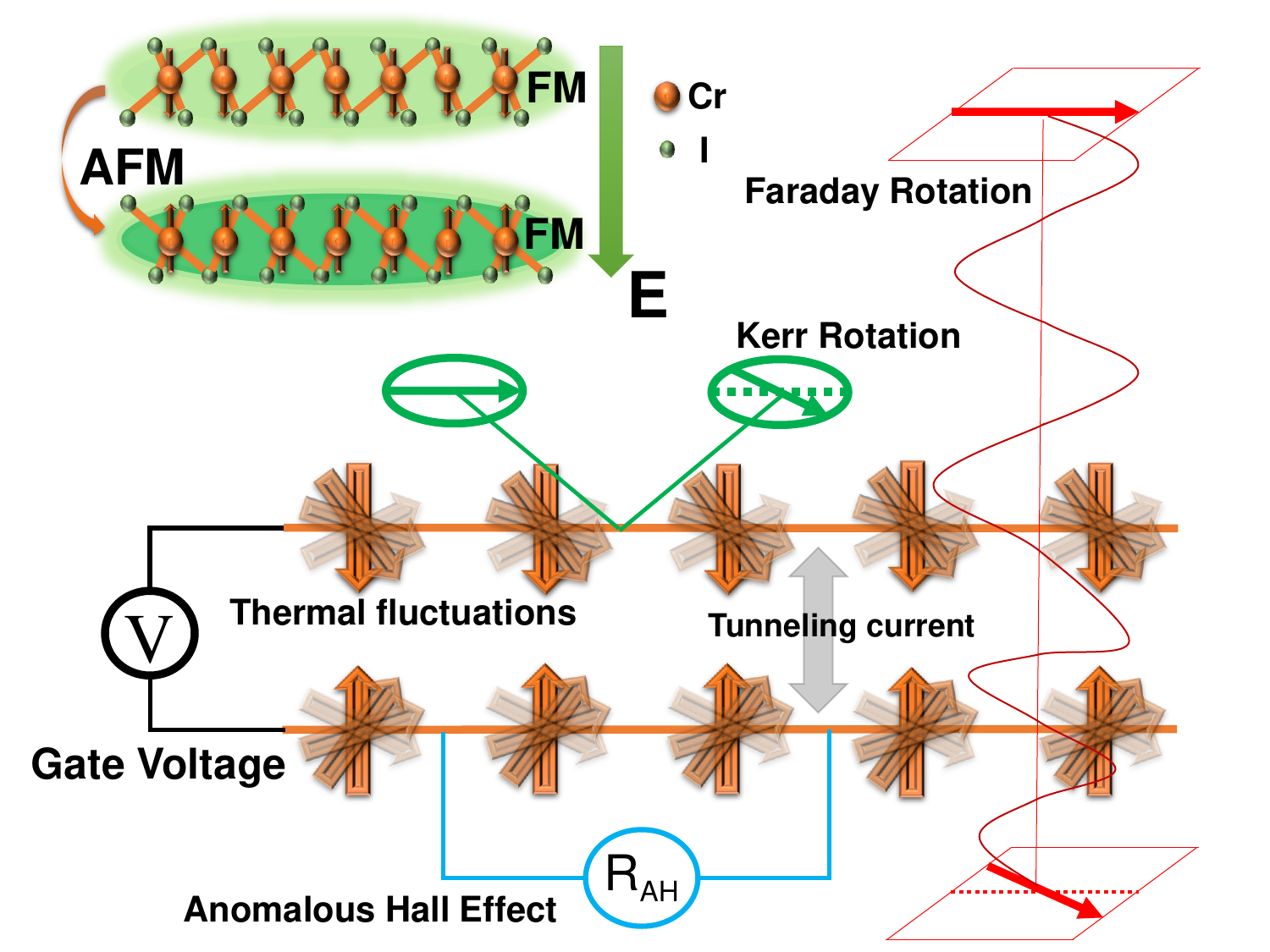} \\
\caption{
Schematic of magnetoelectric response in bilayer $\rm CrI_3$ where
perpendicular gate electric fields polarize the bilayers, making exchange interactions layer dependent.
At finite temperature, this leads to non-zero bulk magnetization, Faraday/Kerr rotation, and anomalous Hall conductivity.
}
\label{me}
\end{figure}

In this paper, we address the possibilities for electrical control of magnetization in CrI$_3$ that exemplifies 
a van der Waals (vdW) semiconductor bilayer with antiferromagnetic interlayer interactions.
We predict by means of Monte Carlo simulations and mean-field calculations %based on the anisotropic Heisenberg model
that antiferromagnetic vdW bilayers at charge neutrality will generally have strong magnetoelectric response when 
the interlayer antiferromagnetic exchange coupling is weaker than intralayer ferromagnetic exchange coupling. We also present a simple model for electrical control of magnetism by electrostatic doping, which has been observed experimentally~ \cite{Jiang2018a,Jiang2018b,Huang2018}.
These effects can be detected in both single-gate devices where we can simultaneously vary the electric fields and carrier densities,
and also in dual-gate geometries where they can be varied independently. 
Our theory presented for CrI$_3$ should find promising applications in other van der Waals antiferromagnetic bilayers
like V-doped WSe$_2$ and CrTe$_2$ ~\cite{roomtemp,CrTe2}
where magnetic ordering at room temperature is expected.
A schematic illustration in Fig.~\ref{me} shows how the magnetoelectric response can be detected 
by measuring the gate voltage dependence of Faraday/Kerr rotation, anomalous Hall conductivity, 
or bulk magnetization.

\section{Theoretical Model for CrI$_3$ Bilayers}
The Heisenberg model for the Cr-ion spin network contains Heisenberg exchange interactions and magnetic anisotropy. The origin of magnetic anisotropy has been widely studied and is still under debate \cite{Lado_2017,Lee2020,Chen2020,Olsen2019,Xu2018,Kim2019,Webster2018,Besbes2019,Pizzochero_2020,Richter2018,Torelli_2018}. It is still not settled whether on-site\cite{Liu2018_DMI,Pizzochero_2020} or exchange anisotropy\cite{Xu2018} dominates in CrI$_3$. However, as we show below, magnetic anisotropy does not play an important role in the magneto-electric response properties, provided it is sufficiently strong to avoid a strong reduction in critical temperature.  We can therefore 
   employ a Heisenberg model with on-site magnetic anisotropy without any loss of generality:
\begin{equation}\label{heisenberg}
% \begin{split}
  {\cal H} = - \frac{1}{2}\sum_{i,\bR,\bRp} {J_{i}(\bR-\bRp)\bS_{i\bR} \cdot \bS_{i\bRp}} - \frac{1}{2} \sum_{\substack{ i\neq j\\ \bR,\bRp}} {J_{v}(\bR-\bRp) \bS_{i\bR} \cdot \bS_{j\bRp}} 
   - D\sum_{i,\bR}{(S_{i\bR}^{z})^2} ~,
% \end{split}
\end{equation}
where $i$ or $j$ are $t,b$ labels for top/bottom layer, $J_i(\bR-\bRp)$ is the intralayer exchange coupling and $J_v(\bR-\bRp)$ the interlayer coupling, $\bR$ is the in-plane Cr ion position, $\bS$ is the spin operator, $D$ is the on-site magnetic anisotropic energy. 
To obtain the parameters of the Heisenberg model for bilayer CrI$_3$,
we have compared the total energies of different spin configurations
using plane wave density-functional-theory as implemented in the Vienna Ab initio simulation package(VASP)~\cite{vasp}
and Quantum Espresso(QE)~\cite{QE} applied to bilayer CrI$_3$, using semi-local PBE-GGA~\cite{pbe} with the
vdW-D2 correction proposed by Grimme~\cite{grimme}.
Both implementations show overall qualitative agreement while minor quantitative differences can exist.
The equilibrium interlayer distances in van der Waals layered materials results from a delicate balance of shorter
ranged chemical bonds and longer ranged Coulomb correlations
which cannot be systematically captured within commonly used local or semi-local density functionals~\cite{vdwinterlayer}.
% --- Addition ---
%
On the other hand, the intralayer electronic structures can rely on the DFT calculations
because of the dominantly covalent or ionic character of the bonds.
We therefore use DFT to evaluate the electric-field dependence of intralayer exchange coupling parameters, while relying on information from experiment for the interlayer exchange interactions obtained 
from the layer AFM-FM crossover induced by a magnetic field. 
The electric field is applied in VASP\cite{Neugebauer1992,Makov1995} by inserting a dipole sheet in the middle of vacuum.
%
% ---
%
Further discussions on the technical details of these calculations can be found in the supporting information~\cite{supplement}.

The anisotropy energy for bilayer $\rm CrI_3$ is $ 660 ~\mu{\rm eV/Cr}$
according to VASP, and is $730~\mu{\rm eV/Cr}$ atom according to QE.
These values are smaller than the corresponding monolayer values ($D=960~\mu{\rm eV/Cr}$ in QE),
and similar to the bulk value $690~\mu{\rm eV/Cr}$~\cite{Zhang2015} reported in the literature.
Magnetic anisotropy plays a critical role in limiting the thermal fluctuations that lower the
monolayer and bilayer critical temperatures relative to their mean-field values.
The increase of $T_c$ with $\Delta$ estimated from Monte Carlo simulations is illustrated in Fig.~S13 ~\cite{supplement}, and is in agreement with results reported in \cite{Torelli_2018}.
Based on  DFT calculations, the magnon energies in bilayer $\rm CrI_3$ are in the range below 15 meV,
and the magnon gap is around 1~meV~\cite{supplement}.
The exchange interaction part of the spin Hamiltonian in Eq.~(\ref{heisenberg}) can
be expressed in momentum-space by using the Fourier transform
$\bS_{i\bR} = 1/\sqrt{N} \sum_{\bR} e^{i \bk \cdot \bR} \bS_{i}(\bk)$ resulting in
\begin{equation}
{\cal H}(\bk) = - \frac{1}{2}\sum_{i,\bk} \, J_{i}(\bk) \, \bS_{i}(-\bk) \cdot \bS_i(\bk) 
- \frac{1}{2}\sum_{\substack{ i\neq j\\ \bk} } J_{v}( \bk ) \, \bS_{i}(\bk) \cdot \bS_{j}(\bk),
\end{equation}
where $i = t/b$ are top/bottom layer labels, $\bS_{i}(\bk)$ is the appropriately normalized
Fourier transform of the dimensionless (without $\hbar$) spin operators in layer $i$, and
\begin{equation}\label{exchange_fourier}
J_{i}(\bk) = \sum_{\bL} \, \exp(-i\bk\cdot \bL) \,  J_{i}(\bL)
\end{equation}
is a Fourier sum of exchange interactions between spins localized at the origin in layer $i$ and at lattice site $\bL \equiv \bR-\bRp$.
%(or $J_{i}(\bL)$ in real-space) with the same indices for
%intralayer exchange constants and $J_{ij}$ with unequal indices for the interlayer exchange constant.
%For a bilayer we denote it as $J_{01}$.
The intralayer exchange constants extracted from ground state energy calculations for different metastable magnetic configurations are summarized in the supporting information~\cite{supplement}.

The overall weakness of interlayer magnetic interaction in CrI$_3$ bilayers can be expected
from the relative weakness of the van der Waals interlayer coupling with respect to the intralayer bonds.
Bulk $\rm CrI_3$ is a layered semiconductor with a low-temperature  R$\bar{3}$ rhombohedral
structure~\cite{McGuire2015,Ding2011,Wang2011a,Feldkemper1998} illustrated in the
supporting information~\cite{supplement} and a high-temperature C2/m monoclinic structure.
We performed calculations using both LDA and LDA+U DFT approximations for both R$\bar{3}$ and C2/m stacking arrangements, with and without spin-orbit coupling.
Within LDA the predicted interlayer magnetic interactions are most often ferromagnetic and much stronger
than experimental estimates, and we find that the monoclinic C2/m structure is ferromagnetic
at equilibrium interlayer distance becoming antiferromagnetic only at larger values of interlayer separation.
The change in sign as $d$ varies is expected, since direct ferromagnetic
exchange interactions are expected to decline more rapidly with $d$ than antiferromagnetic superexchange interactions, and
%
%LDA+U with U$= 3$~eV substantially alters the total energy differences~\cite{Jiang2018}.
additional DFT results are summarized in Fig.~S4 and S5 \cite{supplement}).
Note that the antiferromagnetic and ferromagnetic states reach their minimum energies at different layer separations~\cite{supplement}
and that the vdW gaps of the encapsulated bilayers studied
experimentally~\cite{Huang2018,Jiang2018a,Jiang2018b} are likely smaller than
those of the isolated bilayers we have studied theoretically.
It follows that the field-driven antiferromagnetic to ferromagnetic transition should be accompanied by a change in the separation between layers, and
that the critical field of the transition should be pressure-dependent and altered by encapsulation.

This landscape of DFT results, although not definitively predictive for interlayer exchange, 
%for capturing interlayer exchange interactions, 
establish that the interlayer magnetic interaction in CrI$_3$ bilayers is weak and sensitive to layer separation~\cite{supplement} and stacking arrangement~\cite{supplement,FernandezRossier,Sivadas2018,Jiang2018,Jang2018}.
Below we view the interlayer magnetic interaction parameter as a quantity that is presently most reliably estimated from experiments.
However, as already mentioned, the DFT calculations do reliably predict important details of the intralayer
magnetic interactions $J_{t/b}$ that are related to the covalent intralayer bonding network.

\section{Magnetoelectric Response}
%As discussed above,
The strength of the interlayer exchange constant %$\tilde{J}_{\perp}^{\rm AF} = 
%$ \tilde{J}^{\rm AF}_{01} \equiv {J}_{01} S^2 $
$ \tilde{J}_{v} \equiv {J}_{v} S^2 $ (with $J_v \equiv \sum_{\bL} \, J_{v}(\bL)$)
can be reliably extracted from the magnetic field $B_c \approx 0.65 T$~\cite{Huang2017} needed to drive the antiferromagnetic bilayer to a ferromagnetic state.
This consideration implies that 
$\tilde{J}_{v} 
%$\tilde{J}^{\rm AF}_{01} 
%\tilde{J}_{\perp}^{\rm AF} 
= 2 g \bohr S B_c  \approx - 0.23 \,{\rm meV} $ with S = 3/2.
The interlayer exchange constant is nearly two orders of magnitude
smaller than the bilayer's ferromagnetic intralayer magnetic interaction parameter, which is denoted by $\tilde{J}_{t/b} \equiv J_{t/b} S^2 $,
with $J_{t/b} \equiv \sum_{\bL} \,  J_{t/b}(\bL)$ for top and bottom layers.
In the absence of electric fields we estimate 
that $\tilde{J}_{0} = \tilde{J}_{t} = \tilde{J}_{b} 
%\tilde{J}_{00} = \tilde{J}_{11} 
\approx 14.6$~meV according to DFT calculations.

To establish that weak interlayer coupling implies strong magneto-electric response,
we first apply the mean-field theory
to the intrinsic case where the local spin moments are the only low-energy degree of freedom.
In mean-field theory the temperature dependent moment per site depends only on the exchange coupling at $\bk=0$:
%\begin{eqnarray}
%\langle \bS \rangle_i &=& (-)^i \hat{z} S s_i = (-)^{i} \hat{z} S B_S(\beta h_i) \nonumber \\
%h_i &=& \tilde{J}_{i} s_{i} - |\tilde{J}_{v}| s_{j} + H s_i + (-)^i \lambda \mathcal{E} s_i.
%\label{mft}
%
\begin{eqnarray}
\langle \bS \rangle_{t/b} &=& \pm \hat{z} S s_{t/b} = \pm \hat{z} S B_S(\beta h_{t/b}) \label{mft} \\
%h_{t/b} &=& \tilde{J}_{t/b} s_{t/b} - |\tilde{J}_{v}| s_{b/t} + H s_{t/b} \pm \lambda \mathcal{E} s_{t/b}.
%h_{t/b} &=& \left( \tilde{J}_{0} \pm \lambda \mathcal{E} \right) s_{t/b} - |\tilde{J}_{v}| s_{b/t} + H s_{t/b} .
h_{t/b} &=& \left( \tilde{J}_{t/b} + H \right) s_{t/b} - |\tilde{J}_{v}| s_{b/t}.
\nonumber 
\end{eqnarray}
where the z-axis projected magnetization $s_{t/b}$ at top or bottom layer sites can assume values between $-1$ to 1.  
The $\pm$ signs are associated to top or bottom ($t/b$) layers respectively, 
the Hamiltonian $h_{t/b}$ includes both intra and interlayer exchange coupling. 
The intralayer exchange coupling term $\tilde{J}_{t/b}=\tilde{J}_0 \pm \lambda {\cal E}$ respectively for $t$ and $b$ layers includes an electric field $\mathcal{E}$ dependent  term proportional to the response coefficient $\lambda$, where $ \tilde{J}_0 $ denotes the intralayer exchange coupling in the absence of an electric field. 
%The Zeeman term proportional to 
The magnetic field $H$ is here given in units 
of $\tilde{J}_0$ in the Zeeman term,
%In Eq.~\ref{mft} 
the $\beta = 1/k_BT$ is the inverse temperature,
the intralayer effective exchange coupling $ \tilde{J}_{t/b} \equiv S^2 J_{t/b}(\bk=0) = S^2 \sum_{\bL}{J_{t/b}(\bL)}$ includes all distant neighbor exchange coupling,
and the interlayer $ \tilde{J}_{v} \equiv S^2 J_{v}$. Both terms use S = 3/2 as defined above.
$B_S(x)$ is the Brillouin function, 
and we assume that the magnetization $s_{t/b}$ points in the $\hat{z}$ direction.
In the absence of an electric field, $J_{t}(\bL) = J_{b}(\bL)$ due to inversion symmetry in CrI$_3$ bilayers.
$\tilde{J}_{t/b} > 0$ favors ferromagnetic alignment of spins within each layer,
and $\tilde{J}_{v}<0$, favors antiferromagnetic coupling between layers.
%We thus denote $ \tilde{J}_{AF} \equiv \tilde{J}_{01} $ and $ \tilde{J}_{F} \equiv \tilde{J}_{00} = \tilde{J}_{11} $
%for the interlayer antiferromagnetic and intralayer ferromagnetic coupling parameters.
%in the absent of electric field.

The intralayer magnetic exchange interaction from DFT calculations $\tilde{J}_{t/b} \approx 14.6$~meV
corresponds to a mean-field critical temperature $T_{MF} \sim 94$K.
This size of exchange interaction is in reasonable agreement with experiment
since it would imply a ratio of the critical temperature to its mean-field value $\sim 0.50$,
slightly smaller than the ideal two-dimensional Ising model value $T_c/T_{MF} \sim 0.56$.
The Monte Carlo Curie temperature we obtain for monolayers is 30 K, and the N\'eel temperature for
bilayers is 31 K when we use the experimental antiferromagnetic interlayer coupling $\tilde{J}_{v} \approx -0.23$ meV.
We observe that although $T_c$ is mainly governed by intralayer exchange parameters,
the interlayer coupling can help suppress thermal fluctuations and increase $T_c$.  (See Figs.~S14-S16 \cite{supplement}).

The magnetoelectric coupling we focus on results from mirror-symmetry breaking by a gate tunable electric field ${\cal E}$.
Since $\tilde{J}_{t}+\tilde{J}_{b}$ and $\tilde{J}_{v}$ must be even functions of ${\cal E}$, we can describe linear
magneto-electric response by letting
$\tilde{J}_{t/b}=\tilde{J}_0 \pm \lambda {\cal E}$ as defined earlier. 
%$\tilde{J}_{t/b}=\tilde{J}_0 \pm \lambda {\cal E}$ respectively for $t$ and $b$ layers, where $ \tilde{J}_0 $ denotes the intralayer exchange coupling in the absence of an electric field.
It follows that the linear response to ${\cal E}$ of the total spin per unit cell, where $f=S (s_t-s_b)$ uses the local moments $s_t(s_b)$, \cite{supplement}
\begin{equation} 
\alpha \equiv \frac{df}{d{\cal E}} =  \lambda \rchi_{_F},
\label{eq:alpha}
\end{equation}
where $\rchi_{_F}$ is the magnetic susceptibility and 
\begin{equation}
\rchi_{_F} \equiv \frac{df}{dH} = 2S s_{_{AF}} \frac{\beta B_S^{\prime}(\beta h_0)}{1 -\beta ( \tilde{J}_0-|\tilde{J}_{v}|)  B_S^{\prime}(\beta h_0)},
\end{equation}
where $ B_S^{\prime}(x) $ is the derivative of $ B_S(x)$, $s_{_{AF}} = (s_t+s_b)/2$ is the average local moment of the unit cell in the ${\cal E}=0$ antiferromagnetic bilayer states and becomes 1 and 0 respectively in the layer antiferromagnetic and the layer ferromagnetic configurations in virtue of the layer dependent sign definition in Eq.~(\ref{mft}), see the supporting information~\cite{supplement}.
The $h_0$ corresponds to $h_{t/b}$ with zero electric and magnetic fields.
Since $|\tilde{J}_{v}|$ is always very small compared to $\tilde{J}_0$ in antiferromagnetic vdW bilayers, 
$\tilde{J}_0-|\tilde{J}_{v}| \approx \tilde{J}_0+|\tilde{J}_{v}|$.
Note that $\beta ( \tilde{J}_0 + |\tilde{J}_{v}|)  B_S^{\prime}(\beta h_0)=1$ %(see \cite{supplement}) 
at the antiferromagnetic critical temperature.
The critical divergence in $\rchi$ that would occur at the transition temperature
if the system was ferromagnetic is therefore only weakly truncated in the antiferromagnetic state~\cite{supplement}.
For $T_N/T-1 > |\tilde{J}_{v}|/\tilde{J}_0$ the mean-field magnetoelectric response grows like $(T_N-T)^{-1/2}$.
%This interplay between intralayer ferromagnetism and bilayer antiferromagnetism is a special case of the general relationship between intra-sublattice ferromagnetism and order in bipartite antiferromagnets~\mbox{\cite{smart}}.

\begin{figure}[htp]
\includegraphics[width=0.9\linewidth]{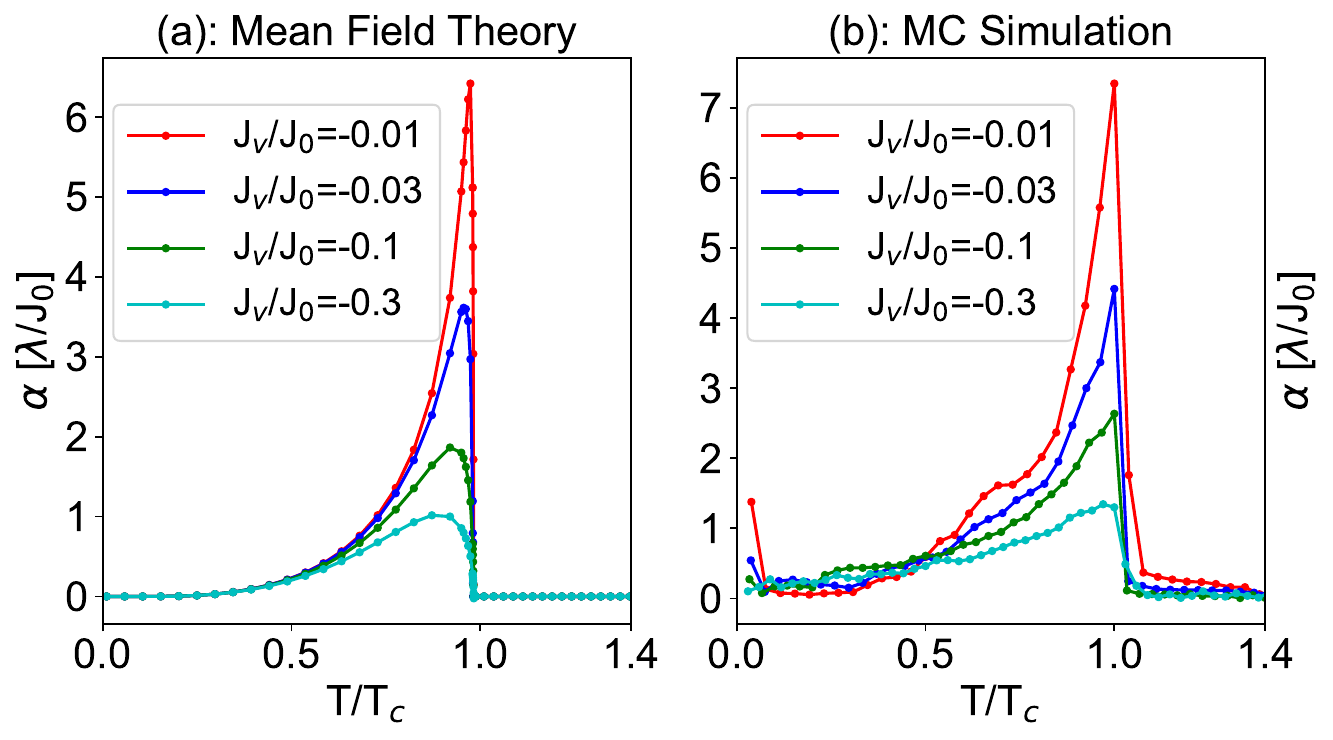} \\
\caption{Magnetoelectric response of antiferromagnetic bilayers in units of $\lambda/J_0$ calculated using (a) mean-field theory and (b) numerical Monte-Carlo (MC) simulation.  The MC simulation was performed using the exchange and magnetic anisotropy of CrI$_3$, except that interlayer exchange was scaled to match the range of values considered in the mean-field calculations.  $J_0$ and $\lambda$ parameters are defined in Eqs.~(\ref{exchange_fourier}), (\ref{lambdadef}) and respectively
characterize intralayer exchange and its sensitivity to external electric fields.
Using DFT estimates for the values of $\lambda$ and $J_{F}$ in CrI$_3$, we find that the ratio of the induced magnetization per volume to the applied electric field is $\sim 10^{-4}$ times the quantity plotted in this figure.
 The MC magneto-electric response 
 $\alpha =  \langle S_z \partial {\cal H}/ \partial {\cal E} \rangle/(N k_BT)$ was calculated from the spin correlation function where ${\cal H}$ is the classical spin-Hamiltonian,
 $S_z$ is proportional to the total magnetization and $N$ is the number of lattice sites per layer.}
\label{figone}
\end{figure}

Fig.~\ref{figone} compares the bilayer magnetoelectric response calculated using this mean-field-theory with the results of classical Monte Carlo simulations  $\alpha = \langle S_z \partial {\cal H}/ \partial {\cal E} \rangle/(N k_BT)$ where ${\cal H}$ is the classical spin-Hamiltonian, $S_z$ is proportional to the total magnetization and $N$ is the number of lattice sites per layer. See the supporting information for further details of the MC calculation. 
From the $\tilde{J}_0 \approx 14.6$~meV DFT results for intralayer exchange interaction 
and the $\tilde{J}_{v} \approx -0.23$~meV experimental results for interlayer exchange interaction,
we have $ {J}_{v}/{J}_0 \approx -0.016$.
Due to the sensitive dependence of interlayer exchange interaction on both interlayer distance and stacking~\cite{supplement},
we calculated the magnetoelectric response for various values of ${J}_{v}/{J}_0$ as shown in Fig.~\ref{figone}.
As we discuss above, the Monte Carlo results are strongly sensitive to magnetic anisotropy
which must be present to endow the mean-field calculations with qualitative validity.
In antiferromagnetic bilayers with the strength of uniaxial anisotropy present in CrI$_3$,
the mean-field predictions are largely validated by Monte Carlo.
The magnetoelectric response is largest
close to but below the antiferromagnetic transition temperature.

In the absence of carriers we find from DFT calculations ~\cite{supplement} that in CrI$_3$ the coupling constant $\lambda$
defined in Eq.~(\ref{eq:alpha}) has the value
\begin{equation}
\label{lambdadef}
|\lambda| = \frac{1}{2} \; \frac{d( \tilde{J}_{t} - \tilde{J}_{b})}{ d {\cal E}} \approx  0.12 \,  {\rm meV}/ \, {\rm V} \,{\rm nm}^{-1}.
\end{equation}
The sign of $\lambda$  is such that the layer with the lower electric
potential and higher charge density has stronger exchange interactions.
Strong gate electric fields ${\cal E} \sim~1~{\rm V}{\rm nm}^{-1}$ yield
intralayer magnetic interactions that are $\sim 2\%$ higher in the high-density
layer than in the low-density layer.
This relatively small variation in the intralayer exchange coupling parameters
can still slightly increase the Curie temperatures ($T_c$) estimated by mean-field and Monte Carlo
calculations~\cite{supplement}.
In the presence of electric field, antisymmetric (Dzyaloshinskii-Moriya exchange interactions (DMI) are allowed by the breaking of inversion symmetry\cite{Liu2018_DMI}.  The DMI increases linearly to $\sim 0.4$ meV for an electric field of 1 V/nm, and thus is smaller than $\sim 0.04$ meV in typical expirimental electric fields up to 0.1 V/nm. This value is small compared with the Heisenberg exchange interactions $\tilde{J}_0 \sim 14.6$ meV and thus plays a negligible role in the physics we address. 
The dependence of interlayer exchange on ${\cal E}$ from DFT calculations is negligible for practical electric field strengths.
Combining the DFT result for $\lambda$ with either the Monte-Carlo or the
mean-field theory results illustrated in Fig. \ref{figone} yields values for the ratio of the
net magnetization per volume to the electric field
$\sim 1 \times10^{-3}$ \footnote{ As ref. \cite{Sivadas2018} pointed out, off-diagonal magneto-electric response coefficients are allowed by symmetry for some stacking arrangements. Based on our DFT calculations, it seems that the off-diagonal response is weak. We place an upper bound as $\sim 5 \times 10^{-5}$ on the off-diagonal response in both R3 and C2m bilayer cases.}  which is dimensionless in cgs units and is  
one order larger than the corresponding values of $10^{-4}$ in
classic magnetoelectric materials like chromia~\cite{Wang2019,Astrov1961,Mostovoy2010}, slightly smaller than that of topological magnetoelectric coefficient (which is around $ 3\times 10^{-3} $).

\section{Influence of Electrostatic Doping CrI$_3$}
The possibility of electrostatic doping expands the phenomenology of magnetoelectric response beyond the effects
of an interlayer potential difference introduced by perpendicular electric fields.
Consideration of carrier doping is experimentally relevant since typical 
samples are often accidentally doped; 
typical carrier densities have been reported to range
from $4.4\times 10^{12} {\rm cm}^{-2}$ \cite{Huang2018} to $2.3\times 10^{13} {\rm cm}^{-2}$\cite{Jiang2018b}.
Both carrier density and electric fields can be controlled independently in dual gated bilayers.

At finite but low carrier densities, the $T=0$ energies of the FM and AF states at a given average electric 
field are given by:
\begin{equation}\label{doping_en}
\begin{aligned}
&    E_{\rm FM} = E_{0}^{\rm FM} - \theta(-n) E_c^{\rm F} n A + \theta(n) E_v^{\rm F} n A
 \\
&    E_{\rm AF} = E_{0}^{\rm AF} - \theta(-n) E_c^{\rm AF} n A + \theta(n) E_v^{\rm AF} n A \\
\end{aligned}
\end{equation}
where $E_{0}^{\rm FM/AF}$ is the energy per 2D unit cell in the absence of carriers,
$n$ is the carrier density per unit cell defined as negative for electrons and positive for holes
defining the value of the Heavyside step function $\theta(n)$,
$A$ is the unit cell area  $\sqrt{3} a^2 / 2$ where $a=6.863$~angstrom %$\AA$
is the lattice constant, and $E_{c/v}$ are the conduction/valence band edge energies.

\begin{figure}[htp]
\includegraphics[width=0.9\linewidth]{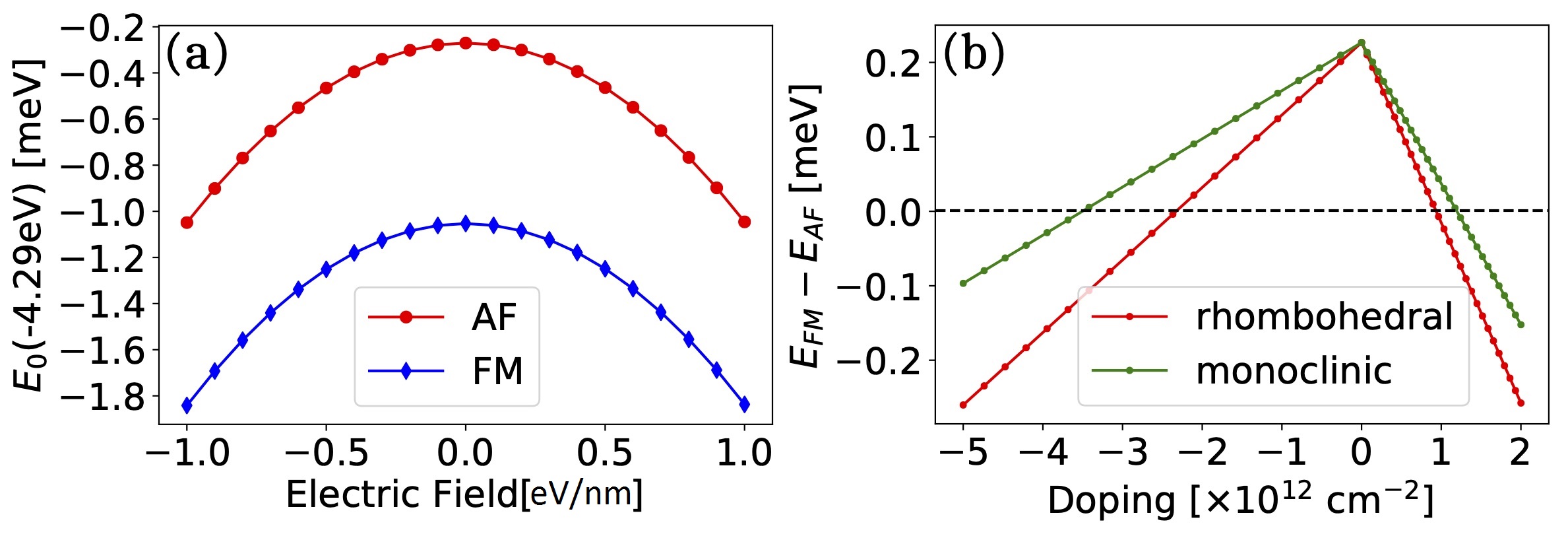}
\caption{Magnetoelectric Effect:
(a) Dependence of ground-state energy on electric field obtained from DFT calculations for the rhombohedrally stacked bilayer.
At zero-carrier density, the ground state energy has a quadratic dependence on the electric field.
%In our {\it ab initio} calculations the FM state is incorrectly lower in energy as discussed in the main text.
(b) Difference between ground state energies as a function of carrier density given in units of electrons
per ${\rm cm^{-2}}$ calculated from Eq.~(\ref{doping_en}) defined as negative (positive) for
electron (hole) doping.
}
\label{magelectricy}
\end{figure}

Doping generally favors ferromagnetic states because the ferromagnetic state has a smaller bandgap,
allowing carriers to be introduced with a smaller band energy cost.
As show in Fig.~\ref{magelectricy}(b), we estimate that carrier densities
of around $1.5 \times 10^{12} {\rm cm}^{-2}$ for electrons and
$3\times 10^{12} {\rm cm}^{-2}$ for holes
are sufficient to induce transitions from antiferromagnetic to ferromagnetic state.
We expect that variations in the total carrier density with gates can change the magnitude of the
interlayer coupling $J_{v}$.  Increasing $J_{v}$ results in magneto-electric effects 
that are weaker at their peak, but sizable over a broader range of temperatures.   

\section{Conclusions}
We have shown that antiferromagnetic van der Waals bilayers 
have a sizable magneto-electric response to gate-applied displacement fields because of their weak interlayer exchange interactions.
The size of the response has a stronger peak, but a lower one away from the critical temperature.
The sign of this response depends on which of the two layers has a particular spin orientation, allowing 
information encoded in the antiferromagnetic bilayer's magnetic configuration 
to be read electrically by measuring the Hall conductivity of the bilayer.
Inducing net magnetism electrically allows, the antiferromagnetic configuration to be 
rewritten by external magnetic fields.  
In dual gated geometries, in which carrier density and displacement field can be varied independently,
the size and temperature dependence of the magneto-electric response can be varied {\em in situ} by 
varying the strength of interlayer interactions.  These properties add to the motivation for 
the development of antiferromagnetic bilayers with robust room temperature order.

%%%%%%%%%%%%%%%%%%%%%%%%%%%%%%%%%%%%%%%%%%%%%%%%%%%%%%%%%%%%%%%%%%%%%
%% The "Acknowledgement" section can be given in all manuscript
%% classes.  This should be given within the "acknowledgement"
%% environment, which will make the correct section or running title.
%%%%%%%%%%%%%%%%%%%%%%%%%%%%%%%%%%%%%%%%%%%%%%%%%%%%%%%%%%%%%%%%%%%%%
\begin{acknowledgement}

C. L. and A. H. M. were supported by the SHINES Center,
funded by the U.S. Department of Energy, Office of Science, Basic Energy Sciences under Award DE-SC0012670,
and the Welch Foundation under grant TBF1473.
B. L. C. was supported by the Basic Science Research Program through the National Research Foundation of Korea (NRF) funded by the Ministry of Education (No. 2018R1A6A1A06024977) and grant NRF-2020R1A5A1016518.
N. B was supported by the grant NRF-2020R1A2C3009142 and 
J. J. by Samsung Science and Technology Foundation under Project No. SSTF-BA1802-06.
K. N. was supported by the JSPS KAKENHI (Grant No. JP15H05854, JP15K21717, JP17K05485) and JST CREST (JPMJCR18T2).
We acknowledge computer time allocations from the Texas Advanced Computing Center and from KISTI through grant KSC-2020-CRE-0072.

\end{acknowledgement}

% \listofchanges

%%%%%%%%%%%%%%%%%%%%%%%%%%%%%%%%%%%%%%%%%%%%%%%%%%%%%%%%%%%%%%%%%%%%%
%% The same is true for Supporting Information, which should use the
%% suppinfo environment.
%%%%%%%%%%%%%%%%%%%%%%%%%%%%%%%%%%%%%%%%%%%%%%%%%%%%%%%%%%%%%%%%%%%%%
\begin{suppinfo}
This material is available free of charge via the internet at http://pubs.acs.org.

\begin{itemize}
 \item Details of cryatal structure, DFT calculations, strain effect,  reliability of interlayer exchange coupling, Monte Carlo simulation, and mean-field theory.
\end{itemize}%

\end{suppinfo}

%%%%%%%%%%%%%%%%%%%%%%%%%%%%%%%%%%%%%%%%%%%%%%%%%%%%%%%%%%%%%%%%%%%%%
%% The appropriate \bibliography command should be placed here.
%% Notice that the class file automatically sets \bibliographystyle
%% and also names the section correctly.
%%%%%%%%%%%%%%%%%%%%%%%%%%%%%%%%%%%%%%%%%%%%%%%%%%%%%%%%%%%%%%%%%%%%%
\bibliography{CrI3}

\end{document}